\documentclass{emulateapj}    
\usepackage{ulem}
\usepackage{color}

\slugcomment{Accepted by ApJ Lett}

\shorttitle{Jet propelled wind ejecta in SS433}
\shortauthors{Blundell \& Hirst}

\begin{document}

\title{Jet propulsion of wind ejecta from a major flare in the black hole microquasar SS433}

\author{Katherine M.\ Blundell\altaffilmark{1} and Paul Hirst\altaffilmark{2}}

\altaffiltext{1}{University of Oxford, Department of Physics, Keble
  Road, Oxford, OX1 3RH, U.K.}
\altaffiltext{2}{Gemini Observatory, 670 North A`oh$\bar{\mathrm o}$k$\bar{\mathrm u}$ Place, University Park, Hilo, HI 96720, U.S.A.}

\begin{abstract}
  We present direct evidence, from Adaptive-Optics near-infra-red
  imaging, of the jets in the Galactic microquasar SS433 interacting
  with enhanced wind-outflow off the accretion disc that surrounds the
  black hole in this system.  Radiant quantities of gas are
  transported significant distances away from the black hole
  approximately perpendicular to the accretion disc from which the
  wind emanates.  We suggest that the material that comprised the
  resulting ``bow-tie'' structure is associated with a major flare
  that the system exhibited ten months prior to the observations.
  During this flare, excess matter was expelled by the accretion disc
  as an enhanced wind, which in turn is ``snow-ploughed'', or
  propelled, out by the much faster jets that move at approximately a
  quarter of the speed of light.  Successive instances of such
  bow-ties may be responsible for the large-scale X-ray cones observed
  across the W50 nebula by ROSAT.
\end{abstract}

\section{Introduction}
Although black holes are popularly thought of as simply sucking matter
towards them, the astrophysical reality is rather different: energetic
jets of matter are ejected in opposite directions away from near
black holes, sometimes emitting as they do so via e.g.\ synchrotron
radiation \citep{Begelman1984}.  Gravitational attraction of matter
occurs simultaneously with angular momentum conservation leading to
matter swirling inwards in a plane and forming structures known as
accretion discs.  As well as jets, these discs themselves give
rise to mass outflow via slow (non-relativistic) winds
\citep{Kuncic2007}, along approximately polar directions to the
instantaneous plane of the accretion disc as it precesses
\citep{Perez2010,BBS08}.  The importance of mass outflow and energy
outflow from black holes is being increasingly recognised as
offsetting cooling and hence playing an important role in influencing
the formation of cosmic structure by quasars and active galaxies but
the details are little known and poorly understood to date
\citep{Bower2008,Cavaliere2008}.  Here we present direct evidence for
the relativistic (0.2c -- 0.3c) jets in the microquasar SS433
propelling gas out to significant distances, in excess of a few
  light months, from its 20\,M$_\odot$ black hole \citep{BBS08}, in a
manner akin to how a snowplough moves snow.  The gas propelled by the
jets originates in the much slower wind from the accretion disc, for
reasons explored in the following sections.

\begin{figure}[htbp]
\begin{center}
\centering
   \includegraphics[width=8.5cm]{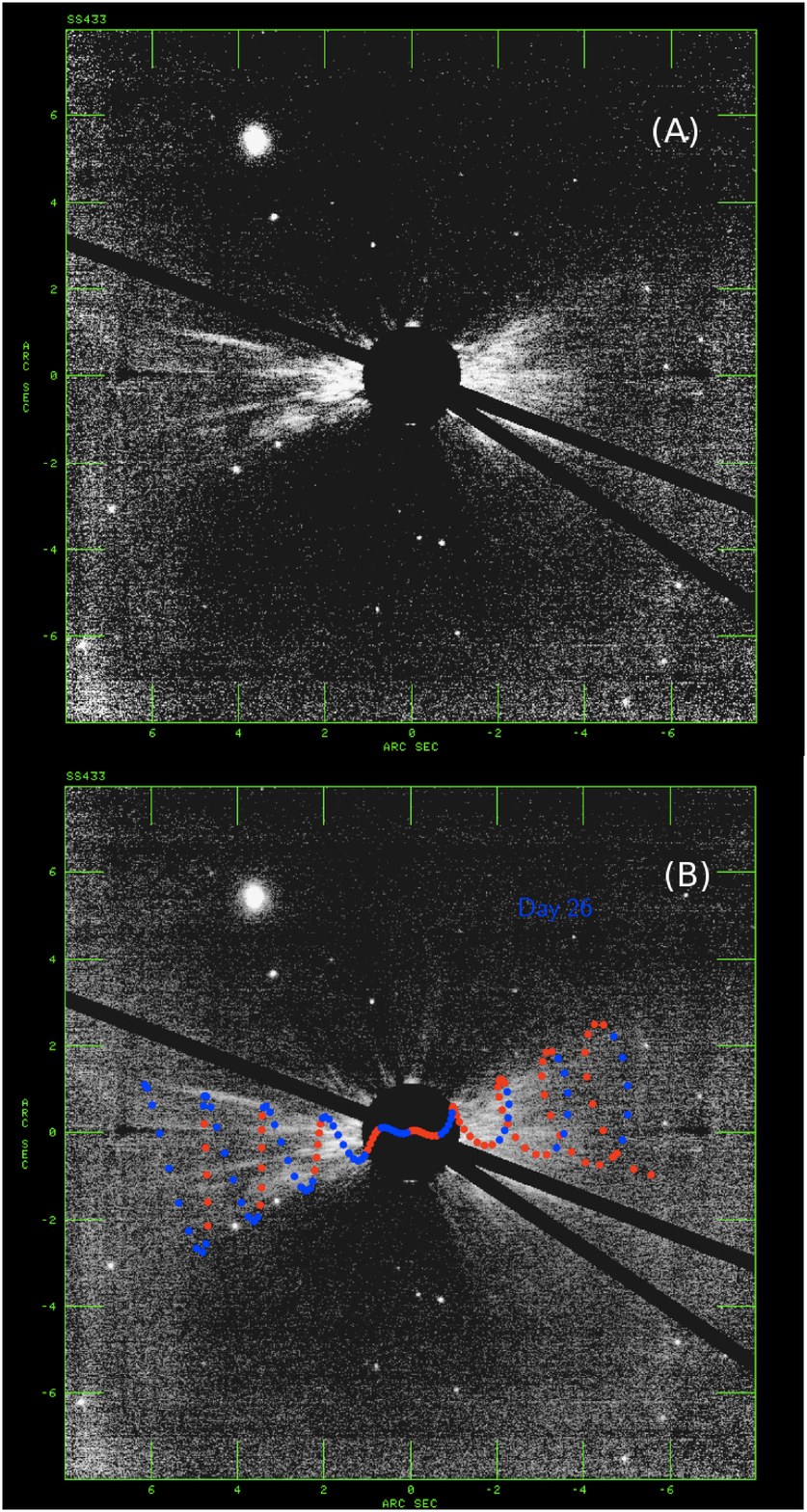} 
   \caption{ Adaptive-optics imaging from VLT-NACO of the microquasar
     SS433 located within our Galaxy at a distance of 5.5 kpc from
     Earth \citep{Blundell2004,Lockman2007}.  Masking in black has
     been applied to the central region (within which azimuthal
     fidelity cannot be relied upon) and also to a diffraction spike
     and filter scratch as described in Sec\,\ref{sec:methods}. The
     lower panel has overlaid coloured dots delineating the location
     of successively launched jet bolides appropriate to the
     precession phase of the infra-red observation (assuming the
       kinematic model is obeyed exactly); those moving towards us
     are coloured blue and those moving away from us are coloured red.
     The correspondence of these with the infra-red emission on the
     east (left) side especially is remarkable.  N.B.\ the two
       point sources on the south edge of the east cone projection are
       believed to be foreground sources, but radial streaks seen
       clearly on the eastern side mimick projected paths that the
       ballastically travelling bollides take.  }
\label{fig:naco}
\end{center}
\end{figure}

\section{Observations and data processing}
\label{sec:methods}
Observations were carried out using the NACO
\citep{Rousset2003,Lenzen2003} instrument at the Nasmyth B focus of
ESO's VLT-UT4 telescope, employing the Ks-filter (centred at
2.2\,$\mu$m which includes the rest-frame Brackett-$\gamma$ at
2.17\,\,$\mu$m) and S13 camera giving a pixel scale of 13\,mas and
field-of-view of 13\,arcsec.  The standard technique of jittering the
telescope position between exposures within a 5-arcsec box centred on
the position of the SS433 nucleus was employed.  Short exposure times
($< 0.6$ seconds) were used to avoid saturating the detector pixels
covering the bright SS433 nucleus. The Adaptive Optics (AO) system was
employed with the visible-light wave-front sensor using the SS433
nucleus itself as the reference object.  The resulting data files were
inspected, and frames with obvious defects (arising from e.g.\ the AO
loop opening, or excessive image elongation) were rejected. 487 frames,
totalling 4287 seconds of exposure time remained.

Both 0.4 and 0.5-second exposure frames were obtained with NACO on
2005-08-18, co-added in groups of 20 by the data acquisition system.
Initial data processing comprised dark correction (using the average
of dark frames taken the following day, and having the same array
configuration and exposure time as the science observations) and
flat-field correction (using the average of the difference between
lamp-on and lamp-off daytime observations of a flat-field calibration
lamp in the same filter and array configuration as the science
frames).  Bad pixels were flagged in both the dark and flat-field
images using a simple threshold to detect unphysical values.
Following the processing of individual data-frames, a novel processing
technique was used to form a clipped-mean stack image.  This
deep-stack image is dominated by the bright central nucleus of SS433,
with the wings of this PSF spreading out over most of the image. To
subtract the PSF, we constructed an azimuthally-averaged radial
profile, centred on the nucleus, and including only those azimuthal
ranges apparently devoid of structure. This radial profile was then
azimuthally swept out to form a circularly-symmetric, model PSF image
that was then subtracted from the deep-stack image data to yield a
PSF-subtracted stacked image.  This PSF subtraction does not model the
detailed (non-circularly symmetric) structure close to the core of the
PSF well, and thus this area has been masked out with 2.2-arcsecond
diameter circular mask. Two other apparent artefacts in the image have
also been masked, believed to result from defects on the filter in use
during the observations.

The individual data frames are dominated by the bright, unresolved
central nucleus of SS433. Imaging artefacts apparent in each frame
include 1) Bad pixels, with data values in either the science or
calibration frames that are inconsistent with being due to incident
light on the array, 2) electronic ghost images of the bright SS433
nucleus, ``reflected'' into the other array quadrants, caused by
electronic cross talk, and 3) diffraction spikes radiating outwards
from the bright SS433 nucleus, due to the presence in the telescope
pupil of the vanes that support the secondary mirror unit of the
telescope. In successive frames, the position angle of these
diffraction spikes rotates in each successive frame with respect to
the sky consistently with the parallactic angle of the observations
(because the azimuth-elevation mount of the telescope means that the
telescope pupil rotates with respect to the sky as the telescope
tracks the target during observations). The next stage of data
reduction was to flag pixels affected by these artefacts as bad so as
to exclude them from subsequent processing. The exact position of the
SS433 nucleus in the image was determined via a 2-d Gaussian fit at
the expected position (inferred on the basis of information recorded
in the header of each image) and exact alignment was then
possible. The cross-talk artefacts were then masked with 50 by 10
pixel bad pixel blocks centred at the SS433 position reflected about
the central pixel of the array.  To mask the diffraction spikes, the
angular offset of the artefacts relative to the telescope parallactic
angle was determined empirically, and two bands of pixels centred on
the SS433 nucleus and 14 pixels wide running across the entire image
at the appropriate angles were flagged as bad.  A clipped-mean stack
from the large number of (short exposure) data frames was performed as
follows: a data-cube was formed from the aligned frames, such that
each ($z$) plane of the cube consists of a single data frame, with the
SS433 nucleus centred in the ($x,y$) plane. The array of values over
$z$ at any given ($x,y$) co-ordinate in the cube thus represents all
our measurements of the flux from that particular point on the sky.

We next looped through each ($x,y$) point, taking the array of values
over $z$ and sorting them by value in the cube, such that extreme
values for any given point on the sky would ``float up'' or ``sink
down'' close to the edges of the cube. We visually inspected the
resulting cube to determine the number of planes to reject at both
ends of the $z$ axis to exclude outliers, rejecting 5 planes from the
top of the cube and 2 from the bottom.  The remaining planes of the
cube were collapsed over the $z$-axis using an arithmetic mean to
generate the deep-stack image. In forming the mean, we ensured pixels
previously masked as bad in the individual frame processing were
excluded.

Fig 1A shows an infra-red image taken with NACO, revealing a
remarkable extended structure that approximately resembles a bow tie
in appearance.  Fig 1B shows that this structures reflects the
projection of the cone traced out by the precession of the launch axis
of its jets \citep{Hjellming1981,Blundell2004}.  Importantly, the
extent (over a few arcseconds) of the bow-tie structure goes way
beyond the regime within a few PSFs (tenths of arcseconds) of the
central star beyond where planet hunting via AO techniques is normally
restricted \citep[e.g.\,][]{Lagrange2009}.  The bow-tie structure does
not resemble any typical AO artefacts either in its azimuthal
dependence or its radial extent \citep[e.g.\,][]{Gladysz2009}.  To
determine if any of the structure could be due to a transient
artefact, the data were re-stacked, split over time into four
equally-sized chunks, and each chunk processed separately through the
stacking and PSF-subtraction methods described.  The bow-tie structure
is clearly visible in each of the separate chunks, indicating that it
is not an artefact correlated with parallactic angle or other
rotational variable.

\begin{figure}[htbp]
\begin{center}
\centering
   \includegraphics[width=8.5cm]{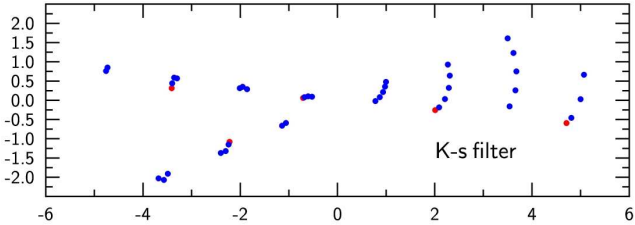} 
%this figure made using /Volumes/Data/projects/ss433_movies/jetprecess_doppler_v3.pl
   \caption{Spectral-filtering figure. This figure shows a prediction
     for the observed distribution of Brackett-$\gamma$ if the
     emitting material is launched in a way that exactly obeys
     the standard kinematic model for the precession of SS433's jet
     axis \citep{Hjellming1981}.    This illustration allows for the fact
     that material moving an inappropriate angle to our line-of-sight
     will be too redshifted or too blueshifted to give
     Brackett-$\gamma$ radiation within the Ks-band filter profile,
     whose lower and upper cut-offs are taken to be 2.05\,$\mu$m
       and 2.27\,$\mu$m.  }
\label{fig:specfilter}
\end{center}
\end{figure}

\section{Emission mechanism}

We now consider what is the mechanism responsible for the bow-tie
emission in the near-IR Ks-band shown in Fig 1, and whether there is
any evidence that the radiating material is moving.  

One possibility
is that the structure is an ionisation cone similar that observed in
LMC-X1 \citep{Cooke2008} but this model would predict a brightness
distribution symmetric on both sides of the nucleus, and marked
east-west asymmetries are observed as examined in
Sec\,\ref{sec:propel}.  Producing the observed structure through the
emission of Bremsstrahlung radiation would require a somewhat
contrived density and temperature profiles, so line emission from the
Brackett-$\gamma$ transition seems to be a more likely explanation. A
definitive demonstration of Brackett-$\gamma$ emission from
fast-moving material would be the detection of specific asymmetric
brightness structures.  For example, if the emitting material were
moving at speed 0.26c radially and ballistically away from the nucleus
along the outside of the precession cone well known from radio imaging
to be traced by the jet axis, and if it were radiating
Brackett-$\gamma$ emission then the Ks-filter's transmission profile
through which the object is observed will act as a velocity filter
because the different angles to the line-of-sight around the cone will
cause the line emission to redshift and blueshift in and out of the
filter profile.  Fig 2 predicts the observed brightness distribution
in this scenario and we note that qualitatively the west wing being
more {\it filled in} and the east wing being more {\it edge
  brightened} reflects the actual observations shown in Fig 1: only
radiation from material that is not too redshifted or too blueshifted
will be observable. We contend that this asymmetric east-west
appearance constitutes circumstantial evidence that the speed with
which the emitting material is moving is close to 0.26c and that the
radiation mechanism is line emission from the Brackett-$\gamma$
transition.

\begin{figure*}[t]
\begin{center}
\centering
   \includegraphics[width=\textwidth]{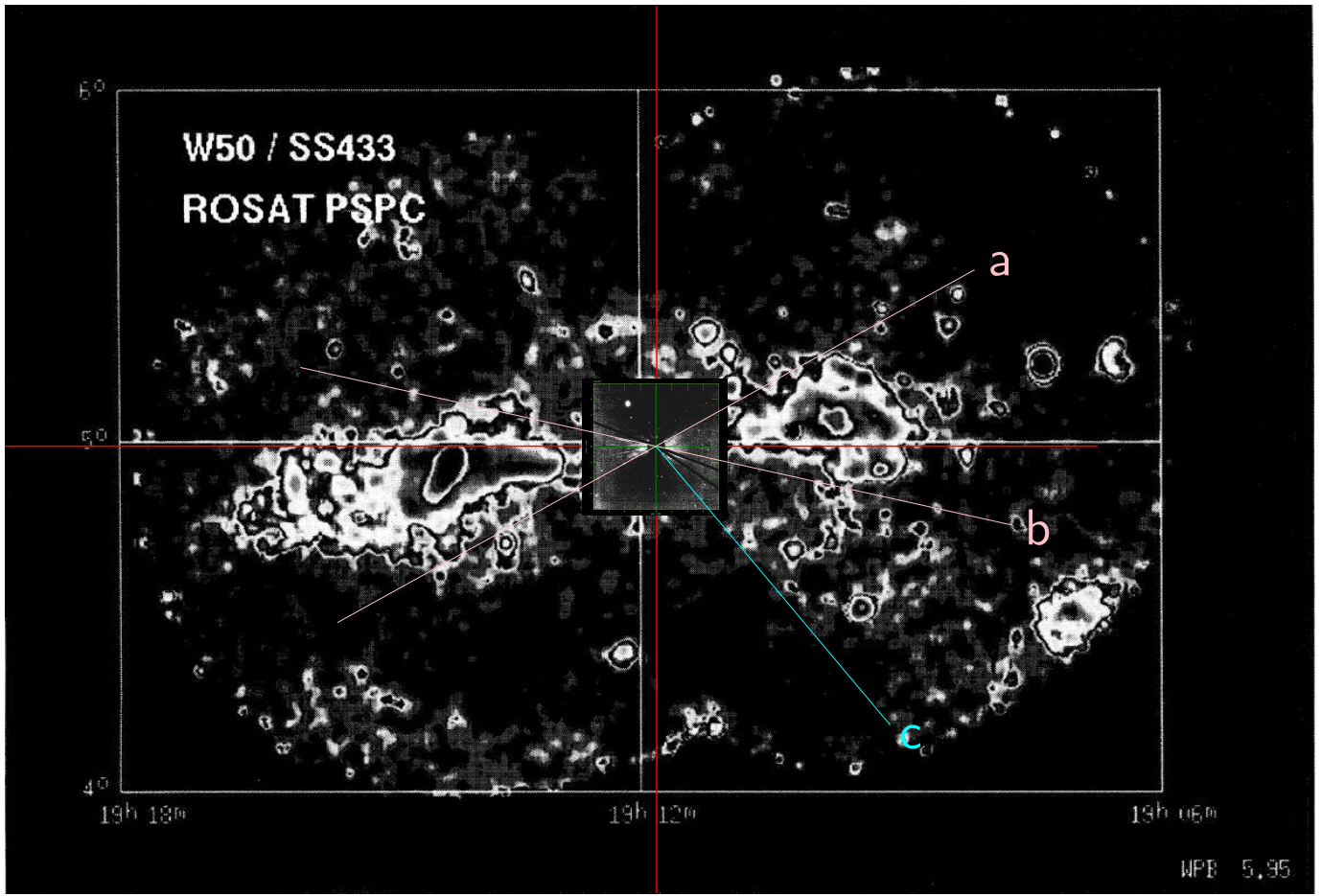} 
   \caption{The background image is X-ray data from the ROSAT
     satellite \citep{Brinkmann1996}; the central inset is the same
     data as Figure 1, enlarged relative to the angular scale of the
     X-ray image.  Pink lines emanate from the central inset that are
     matched to the edges of the eastern part of the infra-red bow
     tie, extended to the western side of the image.  It is
     interesting to note that the bulk of the X-ray emission lies
     centrally within the pink cones.  On the western side we plot in
     cyan a line following the southern edge of the western infra-red
     bow tie.  It is intriguing that this line appears to mark the
     edge of the extended X-ray emission on the lower western side.  }
\label{fig:rosat}
\end{center}
\end{figure*}

\section{Propulsion mechanism}
\label{sec:propel}
If material is travelling ballistically at around the speed of a
quarter of the speed of light (i.e.\ the jet speed), then it will have
left the nucleus at most 10-12 months previously (material travelling
at the sorts of speeds associated with the poloidal disc wind in this
object which is $\sim1000$\,km/s would have travelled only a mere one
third of an arcsec in this time). Ten months before the NACO
observation depicted in Fig 1, i.e.\ on 2004 October 15, SS433's
central engine exhibited a major flare \citep{BST11} involving many
physical phenomena: a spectacular increase in the intensity and speed
of the accretion disc wind and a spectacular increase in the intensity
and launch speed of the jet material, as well as subsequent flaring at
radio wavelengths.  During this flare the intensity of the disc wind,
measured from the stationary hydrogen lines, more than doubled and the
speed of the disc wind more than doubled to 1300\,km/s
\citep{BST11}. The disc wind is known from spectroscopy
\citep{Perez2010,BBS08} to be poloidal and thus approximately aligned
along the cone of the jet axis.  Matter ejected as part of the
enhanced disc wind during that flare, and propelled with the same
range of speeds that the jet material is observed to move at (namely a
range spanning 0.2c to 0.3c --- \citep[see fig\,2c of][]{BB2005}) and
travelling ballistically as the jet material does, will be distributed
along the same radial paths from SS433's nucleus as the emission seen
in Figure 1A, if launched with the same angular distribution as the
jet bolides (Fig 1B).  Note that the stochastic variations in speed
and in direction of the jets \citep{BBS07} will blur out the gas
distribution further.  Thus, remarkably, it would appear that the jets
(well known to move at these speeds) sweep up and propel this
``cloud'' of gas that was ejected as disc wind during the flare, out
to much larger distances by the time of our NACO observation than
would otherwise be the case if moving at speeds of merely $\sim$1000
km/s.  We suggest this process is what gives rise to the radial
streaks (``ionisation trails'') observed most clearly on the left hand
side in Fig 1. There would be no discernible deceleration of the jet
as long as the ram pressure of the jet bolides is sufficiently large
(which translates to the condition that the density through which the
bolides propagate is less than $\sim$10$^{-21}$\,kg \,m$^{-3}$ 
  (assuming the density of the line-emitting bolides inferred by
  \citet{Panferov1997}, their size from \citet{Amy2004}, and expansion
  rate from \citet{BBS07}, and noting that the propagation of the jet
  bolides remains ballastic over many light-months; the simulations of
  \citet{Goodall2011} suggest a density of $3 \times 10^{-22}$\,kg
  \,m$^{-3}$ consistent with this picture).  There was a subsequent
single epoch at which this system was observed with adaptive optics
imaging at Ks-band, 2009 April 24, and at most only hints of a bow-tie
structure were detected on this occasion, although conditions were
remarkably poor for AO observing (1.6$^{\prime\prime}$) seeing.  There
was no major flare that preceded this epoch at an appropriate interval
that would have given rise to the bow-tie structure seen at the first
epoch. We conclude that the bow tie is a transient structure only
present after the central engine of this microquasar has exhibited a
major outburst or flare and expelled significant quantities of
material, and after sufficient time has elapsed (7 -- 10 months) for
the material to be propelled outwards by jet material in its
precessing launch directions.

\section{Relation to other structures \& concluding remarks}
The much larger bow-tie structure observed with ROSAT by
\citet{Brinkmann1996} overlaid in Fig 3 is most likely the
accumulation of lots of major flare output propelled by the mechanism
described in Sec\,\ref{sec:propel}.
Note that the ROSAT bow tie is distinctly different in morphology and
smaller than the 100-pc east-west extent of the W50 conch imaged at
radio wavelengths \citep{Dubner1998}.  The outermost regions of the
W50 radio structure are believed to be caused directly by the
interaction of the radio jets with the supernova remnant seen in the
same radio image.  Line emission takes place within hours of the jet
bolides being launched from the vicinity of the black hole and then
fades rapidly on a timescale of at most a couple of days (from X-ray
spectroscopy \citep{Marshall2008}, optical spectroscopy \citep{BBS07}
and near-IR spectroscopy \citep{Perez2009}).  It is this rapid fading
of the line emission shortly after launch that additionally rules out
``normal'' jet emission (i.e.\ intrinsic to the jet bolides, as
  observed by many authors within a day or so of launch) as an
explanation for the infra-red bow tie shown Fig\,1.  The infra-red
emission most likely arises as line emission from gas heated by the
collision of the expanding bolides with one another as they propagate
away from their launch site, propelling any gas from the vicinity of
the accretion disc that intercepts the path of the wandering,
precessing jet axis.  These results show that the ability to observe
transient behaviour in evolving microquasars can reveal entirely
unexpected astrophysical behaviour.  The serendipitous discovery of
this transient structure is a reminder that the effect of jets of
their surroundings can be considerably more dramatic than observed in
steady state.

\acknowledgments We thank ESO for the
award of DDT-281.D-5062.  We warmly thank Stephen Justham, Robert
Laing, Stephen Blundell, Samuel Doolin, Paul Goodall and Sebastian
Perez for helpful discussions. K.B. thanks the Royal Society for a
University Research Fellowship. P.H. is supported by the Gemini
Observatory, which is operated by the Association of Universities for
Research in Astronomy, Inc., on behalf of the international Gemini
partnership of Argentina, Australia, Brazil, Canada, Chile, the United
Kingdom, and the United States of America.  The referee is gratefully
acknowledged for useful comments on the manuscript.


\begin{thebibliography}{}
\bibitem[Begelman et al.(1984)]{Begelman1984} 
Begelman, M.~C., Blandford, R.~D., \& Rees, M.~J.\ 1984, 
Rev Mod Phys, 56, 255 

\bibitem[Blundell \& Bowler(2004)]{Blundell2004} 
Blundell, K.~M., \& Bowler, M.~G.\ 2004, \apjl, 616, L159 

\bibitem[Blundell \& Bowler(2005)]{BB2005} 
Blundell, K.~M., \& Bowler, M.~G.\ 2005, \apjl, 622, L129 

\bibitem[Blundell, Bowler \& Schmidtobreick(2007)]{BBS07}
Blundell, K. M., Bowler, M. G. and Schmidtobreick, L. 2007
\aap, 474, 903

\bibitem[Blundell et al.(2008)]{BBS08} Blundell, K.~M., 
Bowler, M.~G., \& Schmidtobreick, L.\ 2008, \apjl, 678, L47 

\bibitem[Blundell et al.(2011)]{BST11} Blundell, K.~M., Schmidtobreick
  L., Trushkin S., 2011, accepted by MNRAS and on arXiv.

\bibitem[Bower et al.(2008)]{Bower2008} 
Bower, R.~G., McCarthy, I.~G., \& Benson, A.~J.\ 2008, \mnras, 390, 1399 

\bibitem[Brinkmann et al.(1996)]{Brinkmann1996} 
Brinkmann, W., Aschenbach, B., \& Kawai, N.\ 1996, \aap, 312, 306 

\bibitem[Cavaliere \& Lapi(2008)]{Cavaliere2008} 
Cavaliere, A., \& Lapi, A.\ 2008, \apjl, 673, L5 

\bibitem[Cooke et al.(2008)]{Cooke2008} 
Cooke, R., Bland-Hawthorn, J., Sharp, R., \& Kuncic, Z.\ 2008, \apjl, 687, L29 

\bibitem[Dubner et al.(1998)]{Dubner1998} 
Dubner, G.~M., Holdaway, 
M., Goss, W.~M., \& Mirabel, I.~F.\ 1998, \aj, 116, 1842 

\bibitem[Georganopoulos \& Kazanas(2004)]{Markos2004} 
Georganopoulos, M., \& Kazanas, D.\ 2004, \apjl, 604, L81 

\bibitem[Gladysz \& Christou(2009)]{Gladysz2009} 
Gladysz, S., \& Christou, J.~C.\ 2009, \apj, 698, 28 

\bibitem[Goodall et al.(2011)]{Goodall2011} Goodall, P.~T.,
  Alouani-Bibi, F., \& Blundell, K.~M.\ 2011, MNRAS in press,
  arXiv:1101.3486

\bibitem[Hjellming \& Johnston(1981)]{Hjellming1981} 
Hjellming, R.~M., \& Johnston, K.~J.\ 1981, \apjl, 246, L141 

\bibitem[Junor et al.(1999)]{Junor1999} 
Junor, W., Biretta, J.~A., \& Livio, M.\ 1999, \nat, 401, 891 

\bibitem[King(2010)]{King2010} 
King, A.~R.\ 2010, \mnras, 402, 1516 

\bibitem[Kuncic \& Bicknell(2007)]{Kuncic2007} 
Kuncic, Z., \& Bicknell, G.~V.\ 2007, \apss, 311, 127 

\bibitem[Lagrange et al.(2009)]{Lagrange2009} 
Lagrange, A.-M., et al.\ 2009, \aap, 506, 927 

\bibitem[Laing \& Bridle(2004)]{Laing2004} 
Laing, R.~A., \& Bridle, A.~H.\ 2004, \mnras, 348, 1459 

\bibitem[Lenzen et al.(2003)]{Lenzen2003} 
Lenzen, R., et al.\ 2003, \procspie, 4841, 944 

\bibitem[Lockman et al.(2007)]{Lockman2007} Lockman, F.~J., 
Blundell, K.~M., \& Goss, W.~M.\ 2007, \mnras, 381, 881 

\bibitem[Marshall et al.(2008)]{Marshall2008} 
Marshall, H.~L., Canizares, C.~R., Schulz, N.~S., Heinz, S., Hillwig, T.~C., 
\& Mioduszewski, A.~J.\ 2008, International Journal of Modern Physics D, 17, 1925 

\bibitem[Mioduszewski et al.(2004)]{Amy2004} Mioduszewski, 
A.~J., Rupen, M.~P., Walker, R.~C., Schillemat, K.~M., 
\& Taylor, G.~B.\ 2004, Bulletin of the American Astronomical Society, 36, 967 

\bibitem[Mirabel \& Rodr{\'{\i}}guez(1999)]{Mirabel1999} 
Mirabel, I.~F., \& Rodr{\'{\i}}guez, L.~F.\ 1999, \araa, 37, 409 

\bibitem[Panferov \& Fabrika(1997)]{Panferov1997} 
Panferov, A.~A., \& Fabrika, S.~N.\ 1997, Astronomy Reports, 41, 506 

\bibitem[Perez \& Blundell(2009)]{Perez2009} 
Perez, S., \& Blundell, K.~M.\ 2009, \mnras, 397, 849 

\bibitem[Perez \& Blundell(2010)]{Perez2010} 
Perez, S., \& Blundell, K.~M.\ 2010, \mnras, 408, 2

\bibitem[Rousset et al.(2003)]{Rousset2003} 
Rousset, G., et al.\ 2003, \procspie, 4839, 140 
\end{thebibliography}
\end{document}